\documentclass[10pt]{iopart}
\usepackage{iopams}
\usepackage{color}
\usepackage{graphicx}
\usepackage{natbib}

\def\newblock{\hskip .11em plus .33em minus .07em}

\def\ExB{{\bf E}\times{\bf B}}
\def\tA{\tau_{\rm A}}

\def\tHp{\tau_{\rm Hp}}

\def\SHp{S_{\rm Hp}}
\def\RHp{Re_{\rm Hp}}

\def\qMin{q_{\rm min}}
\def\qRes{q_{\rm s}}

\def\rsi{r_{{\rm s}i}}

\def\rsa{r_{\rm s1}}
\def\rsb{r_{\rm s2}}
\def\rMin{r_{\rm min}}

\def\mMax{m_{\rm max}}
\def\mPeak{m_{\rm peak}}

\def\flPsi{\widetilde{\psi}}

\def\eqPhi{\overline{\phi}}
\def\eqVor{\overline{u}}
\def\eqCur{\overline{j}}
\def\eqF{\overline{F}}

\def\gLin{\gamma_{\rm lin}}

\def\Ma{M^{(1)}}
\def\Mb{M^{(2)}}

\def\O{\mathcal{O}}

\definecolor{gray}{rgb}{0.5,0.5,0.5}
\definecolor{dred}{rgb}{0.5,0.0,0.0}
\definecolor{dgreen}{rgb}{0.0,0.5,0.0}
\definecolor{dblue}{rgb}{0.0,0.0,0.5}

\begin{document}

\title[Resistive and collisionless double tearing modes]{Comparison between resistive and collisionless double tearing modes for nearby resonant surfaces}

\author{A Bierwage\footnote{Present address: Department of Physics and Astronomy, University of California, Irvine, CA 92697} and Q Yu}
\address{Max-Planck-Institut f\"{u}r Plasmaphysik, EURATOM Association, \\ D-85748 Garching, Germany}
\eads{\mailto{abierwag@uci.edu}, \mailto{qiy@ipp.mpg.de}}

\begin{abstract}
The linear instability and nonlinear dynamics of collisional (resistive) and collisionless (due to electron inertia) double tearing modes (DTMs) are compared with the use of a reduced cylindrical model of a tokamak plasma. We focus on cases where two $q=2$ resonant surfaces are located a small distance apart. It is found that regardless of the magnetic reconnection mechanism, resistivity or electron inertia, the fastest growing linear eigenmodes may have high poloidal mode numbers $m \sim 10$. The spectrum of unstable modes tends to be broader in the collisionless case. In the nonlinear regime, it is shown that in both cases fast growing high-$m$ DTMs lead to an annular collapse involving small magnetic island structures. In addition, collisionless DTMs exhibit multiple reconnection cycles due to reversibility of collisionless reconnection and strong $\ExB$ flows. Collisionless reconnection leads to a saturated stable state, while in the collisional case resistive decay keeps the system weakly dynamic by driving it back towards the unstable equilibrium maintained by a source term.
\end{abstract}


\section{Introduction}

Non-monotonic current density profiles, where the maximum current density is located off the magnetic axis, are frequently produced in tokamak plasmas (see \cite{Bierwage05b} and references therein). These so-called reversed-shear (RS) configurations are of considerable interest for establishing high-performance discharges with improved confinement (e.g., \cite{Kikuchi93, Goldston94, Connor04a}). The non-monotonic current profile is associated with a safety factor profile $q(r)$ that has a minimum $\qMin$ at some radius $\rMin > 0$. Around $\rMin$, pairs of magnetic surfaces where $q$ has the same rational value $\qRes = m/n$ can occur a small distance $D_{12}$ apart. Under such conditions, coupled resonant perturbations (with poloidal mode number $m$ and toroidal mode number $n$) known as double tearing modes (DTMs) can become unstable \cite{Furth73, Pritchett80}.

The DTM is a stronger instability than an ordinary tearing mode \cite{Furth63} and bears similarity with the $m=1$ internal kink mode \cite{Pritchett80, Coppi76}. Several nonlinear studies of cases with relatively large inter-resonance distances and dominant low-$m$ modes were conducted in the past (e.g.,~\cite{White77, Persson94, Yu96, Ishii02}). It has recently been shown that DTMs with high poloidal mode numbers $m\sim 10$ may become strongly unstable when the distance $D_{12}$ between the resonances is small \cite{Bierwage05a}. The linear instability of resistive DTMs in such cases was analyzed in detail in \cite{Bierwage05b}.

The present work is motivated by the question how the linear instability and nonlinear evolution of DTMs in configurations with small inter-resonance distance $D_{12}$ depend on the reconnection mechanism, and what role high-$m$ modes play. We approach this question by comparing the dynamics of collisional and collisionless DTMs where magnetic reconnection is mediated by resistivity and electron inertia, respectively. The practical motivation for this work lies in the fact that scenarios with small distance $D_{12}$ inevitably occur during the evolution of the $q$ profile when $\qMin$ passes through low-order rational values $\qRes$. Moreover, in tokamak plasmas of interest to thermonuclear fusion applications the classical resistivity is low, so models which include a collisionless reconnection mechanism may give a more realistic picture. Note that the attribute ``collisionless'' refers to the bulk of the plasma, whereas sufficiently peaked current sheets eventually experience dissipation, e.g., due to ``anomalous'' resistivity \cite{Ji98, Numata02} or electron viscosity \cite{Kaw79, Aydemir90, Yu95, Dong03}. The results may be useful for understanding magnetohydrodynamic (MHD) activity observed near $\qMin$ in RS tokamak configurations \cite{Levinton98, Guenter00} and may bear relevance to problems of stability, confinement and current profile control.

Due to similarities between strongly coupled DTMs and $m=1$ internal kink modes the present work is related to previous studies on fast collisionless reconnection, some of which used a model similar to the reduced set of MHD equations employed here (e.g.,~\cite{Wesson90, Drake91, BiskampDrake94, Ottaviani95, Ishizawa03}). For the sake of simplicity and transparency, several potentially important physical effects (e.g., finite-Larmor-radius corrections and diamagnetic drifts \cite{Coppi64b, Rogers96, Yu03}) are ignored at the present stage.

In the first part of this paper, it is shown that collisionless DTMs may also have a broad spectrum with dominant high-$m$ modes when the inter-resonance distance is small, so they are similar in this respect to resistive DTMs. When resistivity or the electron skin depth are increased, the mode number of the fastest growing mode $\mPeak$ increases. A significant difference between the two cases is that the width of the spectrum of unstable DTMs increases with increasing electron skin depth, whereas resistive DTMs tend to have a fixed spectral width independent of resistivity.

In the second part, nonlinear simulation results are presented. Both cases, resistive and collisionless, have in common an annular collapse involving small magnetic islands structures. In addition, collisionless reconnection converts magnetic energy into kinetic energy more efficiently, which results in strong $\ExB$ flows. This and the reversibility inherent to collisionless reconnection \cite{Ottaviani95} leads to multiple reconnection cycles. Secondary reconnection was previously demonstrated for the $m=1$ internal kink mode \cite{BiskampDrake94} and is here shown to occur in similar form with DTMs. It is essentially an overshoot phenomenon and thus much more pronounced in systems where dissipation is weak.

This paper is organized as follows. In section~\ref{sec:model} the physical model is introduced and section~\ref{sec:numerics} contains details of the numerical methods employed. In section~\ref{sec:equlib-lin} we describe the equilibrium configuration used and its linear instability characteristics. Nonlinear simulation results are presented in section~\ref{sec:results}, followed by a discussion and conclusions in section~\ref{sec:conclusion}.

\section{Model}
\label{sec:model}

We use a reduced set of magnetohydrodynamic (RMHD) equations in cylindrical geometry in the limit of zero pressure \cite{Strauss76, NishikawaWakatani}. The RMHD model has proven to be useful in studies of MHD instabilities when the focus is on a qualitative description of fundamental aspects of the magnetized plasma system, as is the case here. We use an Ohm's law that includes electrical resistivity, electron inertia and perpendicular electron viscosity,
\begin{equation}
{\bf E} + {\bf v}\times{\bf B} = \eta J + \frac{m_e}{n_e e^2}\frac{{\rm d}J}{{\rm d}t} + \frac{\mu_{\rm e}}{\epsilon_0\omega_{\rm pe}^2} \nabla_\perp^2 J,
\label{eq:ohm}
\end{equation}

\noindent where $\eta$ is the resistivity, $\mu_{\rm e}$ the perpendicular electron viscosity, $n_{\rm e}$ the electron density, $m_{\rm e}$ the electron mass, and $\omega_{\rm pe} = \sqrt{n_{\rm e} e^2 / (\epsilon_0 m_{\rm e})}$ the electron plasma frequency. The RMHD equations govern the evolution of the generalized flux function $F$ and the electrostatic potential $\phi$. They are, in normalized form,
\begin{eqnarray}
\partial_t F &=& \left[F,\phi\right] - \partial_\zeta\phi + \SHp^{-1}\left(\hat{\eta}\nabla_\perp^2 F - E_0\right),
\label{eq:rmhd1}
\\
\partial_t u &=& \left[u,\phi\right] + \left[j,\psi\right] + \partial_\zeta j + \RHp^{-1}\nabla_\perp^2 u.
\label{eq:rmhd2}
\end{eqnarray}

\noindent Here, $F$ is defined in terms of the magnetic flux $\psi$ and current density $j$ as $F \equiv \psi + d_{\rm e}^2 j$, with $d_{\rm e} = \sqrt{m_e /(\mu_0 n_e e^2)}$ being the collisionless electron skin depth. The time is measured in units of the poloidal Alfv\'{e}n time, $\tHp = \sqrt{\mu_0 \rho_{\rm m}} a/B_0$, and the radial coordinate is normalized by the minor radius $a$ of the plasma. $\rho_{\rm m}$ is the mass density and $B_0$ the strong axial magnetic field. The current density $j$ and the vorticity $u$ are related to $\psi$ and $\phi$ through $j = -\nabla_\perp^2\psi$ and $u = \nabla_\perp^2\phi$, respectively. 

The strength of the diffusion term in equation~(\ref{eq:rmhd1}) is measured by the magnetic Reynolds number $\SHp = \tau_\eta / \tHp$, with $\tau_\eta = a^2\mu_0/\eta_0$ being the resistive diffusion time and $\eta_0=\eta(r=0)$ the electrical resistivity in the plasma core. This term has two components, $\SHp^{-1}\nabla_\perp^2 F = -\SHp^{-1} j + d_e^2 \mu_{\rm e}\nabla_\perp^2 j$, which are due to the electrical resistivity and the perpendicular electron viscosity, respectively. This convenient form requires that $\mu_{\rm e}$ has the same value as $\SHp^{-1}$ (so $d_{\rm e}^2\mu_{\rm e} \ll \SHp^{-1}$), although these two parameters are physically independent.
In our nonlinear simulations of the collisionless case the magnitude of the electron viscosity term is often measured to be about one order of magnitude larger than the resistive term due to the higher-order derivative. Flow damping at small scales is provided by an ion viscosity term in equation~(\ref{eq:rmhd2}). Its strength is determined by the kinematic Reynolds number $\RHp = a^2/\nu\tHp$, where $\nu$ is the perpendicular ion viscosity.

The source term $\SHp^{-1}E_0$ in equation~(\ref{eq:rmhd1}), with $E_0 = \hat{\eta}\eqF$, balances the resistive diffusion of the equilibrium current profile $\eqCur(r)$. In nonlinear calculations for the collisional case the resistivity profile is given in terms of the equilibrium current density distribution as $\hat{\eta}(r) = \eqCur(r=0)/\eqCur(r)$ (constant loop voltage, $E_0 = {\rm const}$). For simplicity, the temporal variation of the resistivity profile $\hat{\eta}$ is neglected. The effect of $\SHp E_0$ is negligible in the collisionless case, where $\hat{\eta} = 1$ is used.

Each field variable $f$ is decomposed into an equilibrium part $\overline{f}$ and a perturbation $\widetilde{f}$ as
\begin{equation}
f(r,\vartheta,\zeta,t) = \overline{f}(r) + \widetilde{f}(r, \vartheta, \zeta, t).
\end{equation}

\noindent The system is described in terms of the Fourier modes, $\psi_{m,n}$ and $\phi_{m,n}$, obtained from the expansion
\begin{equation}
f(r, \vartheta, \zeta, t) = \frac{1}{2}\sum_{m,n} f_{m,n}(r,t)\; e^{i(m\vartheta - n\zeta)} + {\rm c.c.},
\end{equation}

\noindent with $m$ being the poloidal mode number and $n$ the toroidal mode number. The $(m,n)$ subscripts are often omitted for convenience. We consider only the dynamics within a given helicity $h=m/n={\rm const}$, so the problem is reduced to two dimensions.

\section{Numerical method}
\label{sec:numerics}

For the numerical solution of the model equations (\ref{eq:rmhd1}) and (\ref{eq:rmhd2}) a two-step predictor-corrector method is applied. In the first time step, the dissipation terms are treated implicitly, all others explicitly, and the field variables are estimated at an intermediate time step $t+\Delta t/2$. The second is a full time step, $t \rightarrow t+\Delta t$, with the right-hand sides of equations~(\ref{eq:rmhd1}) and (\ref{eq:rmhd2}) evaluated at the intermediate time step $t+\Delta t/2$ estimated before. In the nonlinear regime, the time step size is of the order $\Delta t \sim 10^{-3}$.

Up to 128 Fourier modes (including $m=0$) are carried, while Poisson brackets $[f,g] = \frac{1}{r}(\partial_r f \partial_\vartheta g - \partial_r g \partial_\vartheta f)$ are evaluated in real space (pseudo-spectral technique, dealiased). The radial coordinate is discretized with a non-uniformly spaced grid, with a grid density of up to $N_r^{-1} = 1/6000$ in regions where sharp current density peaks occur. A fourth-order centered finite-difference method is applied for the $\partial_r$-terms in the Poisson brackets. The Laplacians $\nabla_{\perp(m,n)}^2 = \frac{1}{r}\partial_r r\partial_r - m^2/r^2$ are evaluated at second-order accuracy (tridiagonal matrix equations).

Periodic boundary conditions are applied in the azimuthal and axial directions. At $r=1$ an, ideally conducting wall is assumed, requiring all perturbations to be identical to zero at that location: $\widetilde{f}(r=1) = 0$ (fixed boundary, no vacuum region). At $r=0$, additional boundary conditions are applied to ensure smoothness: $\partial_r \widetilde{f}_{m=0}(r=0) = 0$ and $\widetilde{f}_{m \neq 0}(r=0) = 0$.

The linear dispersion relations and mode structures presented in the following section were computed with both an initial-value-problem (IVP) solver (linearized version of the numerical code described above) and an eigenvalue-problem (EVP) solver \cite{Bierwage05b}. The results of both approaches agree. Results obtained with the EVP solver which the IVP solver cannot produce [such as multiple eigenmodes for given $(m,n)$] were verified by checking the numerical convergence with increasing grid density.

\begin{figure}
[tb]
\centering
\includegraphics[height=6.0cm,width=8.0cm]
{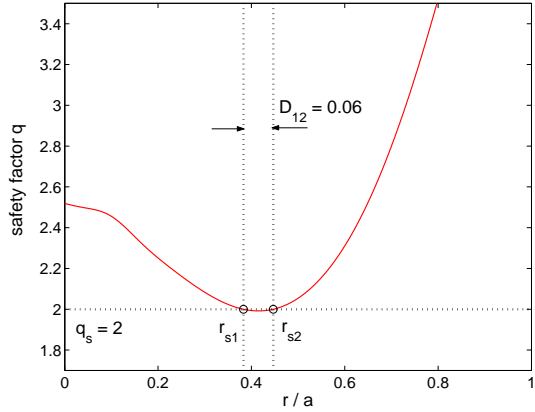}
\caption{Equilibrium safety factor profile $q(r)$. Two $\qRes = 2$ resonant surfaces $\rsa$ and $\rsb$, indicated by vertical dotted lines, are located a small distance $D_{12} = 0.06$ apart. This $q$ profile can be reproduced with the model formula (11) in \protect\cite{Bierwage05b}, with the parameter values of case~(IIIb) in that reference.}
\label{fig:equlib_q}%
\end{figure}

\section{Equilibrium and linear instability}
\label{sec:equlib-lin}

The equilibrium state is taken to be axisymmetric (only $m=n=0$ components) and free of flows, i.e., $\eqPhi = \eqVor = 0$. The equilibrium magnetic configuration is uniquely defined in terms of the safety factor $q(r)$. The magnetic flux function and current density profiles are given by the relations
\begin{equation}
q^{-1} = -\frac{1}{r} \frac{{\rm d}}{{\rm d}r}\psi_{0,0} \quad {\rm and} \quad j_{0,0} = \frac{1}{r} \frac{{\rm d}}{{\rm d}r} \frac{r^2}{q}.
\label{eq:q-equlib}
\end{equation}

\noindent The form of the $q$ profile is shown in figure~\ref{fig:equlib_q}. The two resonant surfaces considered are $\qRes \equiv q(\rsi) = 2$ ($i=1,2$). Their distance is $D_{12} = |\rsb - \rsa| = 0.06$ and the values of the magnetic shear $s=rq'/q$ at the resonances are $s_1 = -0.10$ and $s_2 = 0.12$.

\begin{figure}
[tb]
\centering
\includegraphics[height=6.0cm,width=8.0cm]
{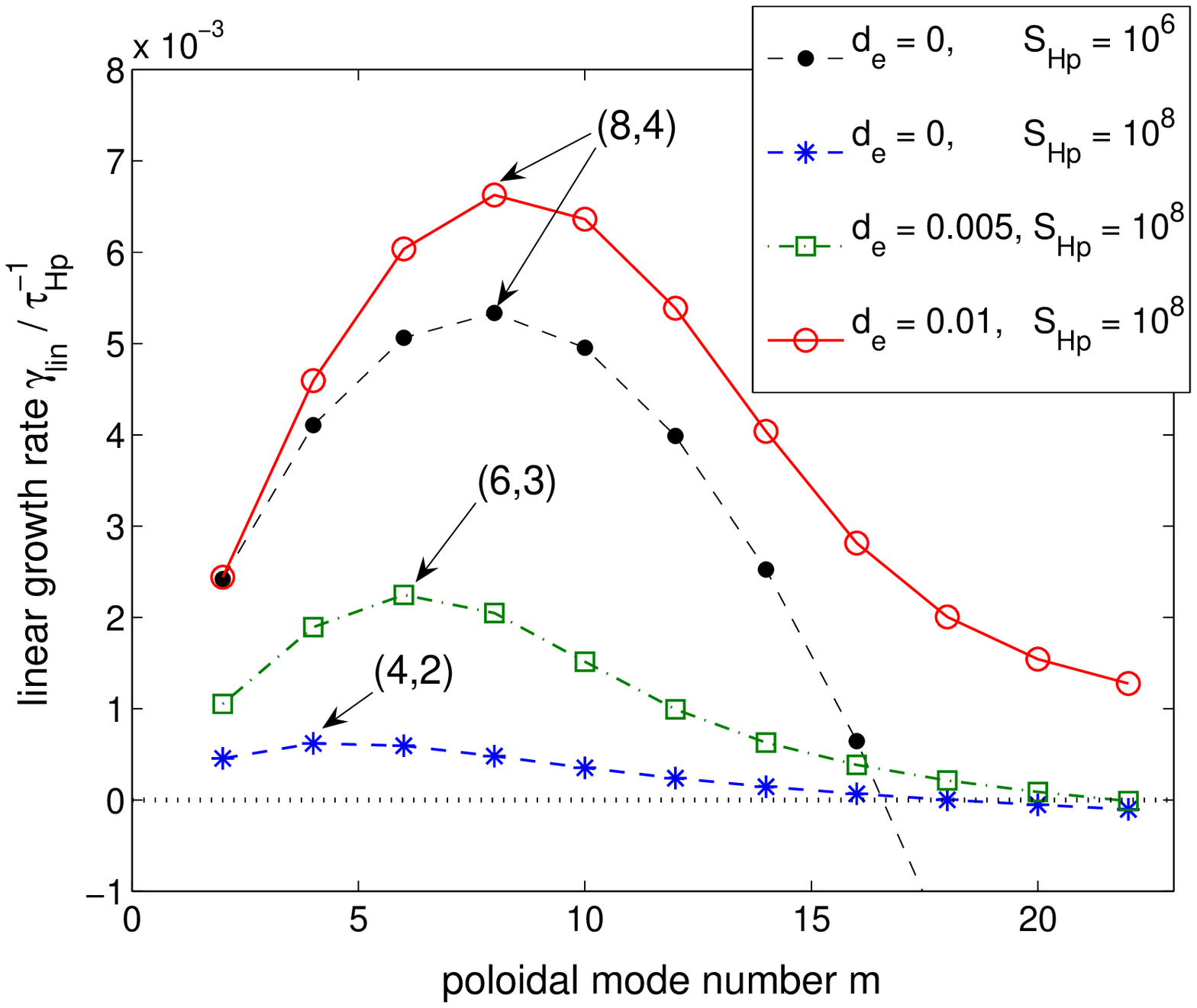}
\caption{Growth rate spectra $\gLin(m)$ of unstable DTM eigenmodes for the $q$ profile in figure~\protect\ref{fig:equlib_q}. (\fullcircle): collisional case studied in this paper ($\SHp = 10^6$, $\RHp = 10^7$, $d_{\rm e} = 0$). For the parameter values $\SHp = 10^8$ and $\RHp = 10^7$, further spectra are shown for $d_{\rm e} = 0$, $0.005$ and $0.01$. The case with $d_{\rm e} = 0.01$ (\opencircle) is the one used in this paper to study the nonlinear evolution of collisionless DTMs. Only growth rates on the dominant eigenmode branch ($\Mb$-type, cf.~figure~\protect\ref{fig:mstruc}) are shown. The fastest growing modes are indicated by arrows.}
\label{fig:spec_dtm}%
\end{figure}

\begin{figure}
[tb]
\centering
\includegraphics[height=6.0cm,width=8.0cm]
{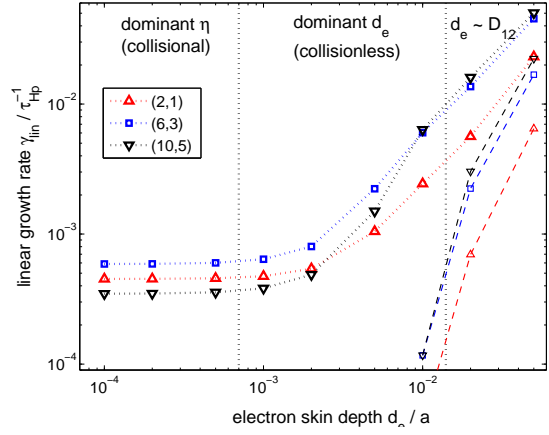}
\caption{$d_{\rm e}$ dependence of the linear growth rate of the modes $m=2$, $6$ and $10$. The scanned range $10^{-4} \leq d_{\rm e} \leq 5\times 10^{-2}$ is roughly divided into three regimes: predominantly collisional, collisionless, and a regime where the skin depth $d_{\rm e}$ becomes comparable to the inter-resonance distance $D_{12}$. Both eigenmode branches $\Ma$ (\broken) and $\Mb$ (\dotted) are shown (cf.~figure~\protect\ref{fig:mstruc}).}
\label{fig:scan-de}%
\end{figure}

\begin{figure}
[tb]
\centering
\includegraphics[width=13.0cm]
{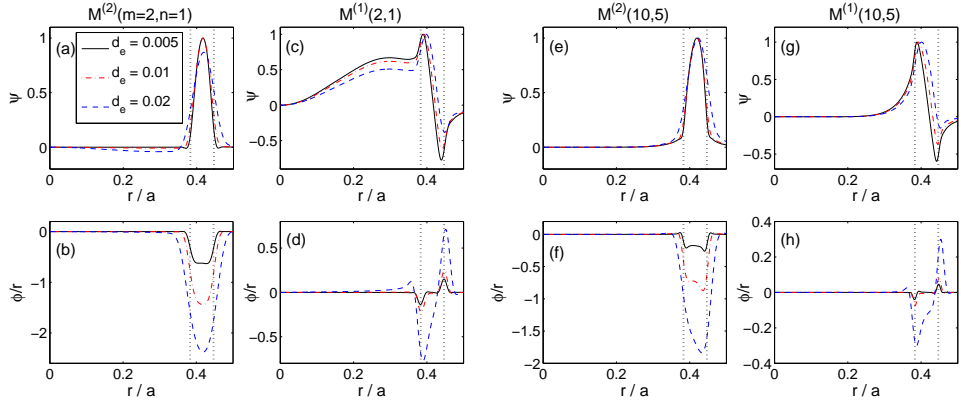}
\caption{Eigenmode structures of collisionless modes with $m=2$ and $10$ in dependence of $d_{\rm e}$. The eigenmode of type $\Mb$ (a,b,e,f) is unstable in the whole range of $d_{\rm e}$ shown in figure~\protect\ref{fig:scan-de} and has odd parity. $\Ma$-type modes (c,d,g,h) have even parity and are destabilized when $d_{\rm e}$ becomes comparable to $D_{12}$.}
\label{fig:mstruc}%
\end{figure}

The linear dispersion relation (spectrum of linear growth rates) $\gLin(m)$ is plotted in figure~\ref{fig:spec_dtm} for collisional and collisionless cases. Increasing the electron skin depth $d_{\rm e}$ increases the linear growth rates, as is to be expected. In addition, an increase in the mode number of the fastest growing mode, $\mPeak$, is observed. The results in figure~\ref{fig:spec_dtm} show that the dominance of modes with $m > 2$ is a feature common to both collisional \cite{Bierwage05b} and collisionless DTMs when the distance $D_{12}$ is small.

A remarkable difference between collisional and collisionless DTMs is that $\mMax$, the mode number of the last unstable mode [$\gLin(m) > 0$ for $m \leq \mMax$] increases with increasing $d_{\rm e}$, as can be seen in figure~\ref{fig:spec_dtm}. In the case of collisional DTMs a variation of $\SHp$ does not affect $\mMax$ (here, $\mMax = 16$) (cf.~also \cite{Bierwage05b}). This property has the important implication that the instability of a DTM with a given mode number $m$ is not only determined by the global current profile. Further calculations have shown that setting the electron viscosity $\mu_{\rm e}$ to zero reduces the growth rates in the high-$m$ domain, but it does not remove the characteristic high-$m$ tail of the collisionless DTM spectrum. This observation indicates that details of the mode structure near the resonant surfaces may also play a role, which requires further investigation.

The $d_{\rm e}$ dependence of the growth rates of individual modes, $(m,n) = (2,1)$, $(6,3)$ and $(10,5)$, is shown in figure~\ref{fig:scan-de} for $\SHp = 10^8$ and $\RHp = 10^7$. The collisional regime is $d_{\rm e} \lesssim 7\times 10^{-4}$. Here the electron inertia plays no significant role. In the range $10^{-3} < d_{\rm e} \lesssim 10^{-2}$ we speak of collisionless DTMs. Here the growth rates rise steeply with $d_{\rm e}$, and the $m=10$ mode undergoes the strongest destabilization among the modes plotted. Finally, for $d_{\rm e} > 10^{-2}$ the skin depth becomes comparable to the inter-resonance distance $D_{12}$, \textit{i.e.}, there is no ideal-MHD layer between the resonant surfaces (for a theoretical study of this regime see \cite{Mahajan82}).
In this regime, a second unstable eigenmode arises for each $(m,n)$ (small symbols connected by broken lines in figure~\ref{fig:scan-de}).

The eigenmode structures for collisionless modes with $(m,n) = (2,1)$ and $(10,5)$ are shown in figure~\ref{fig:mstruc}. The $\Mb$-type mode is the dominant one in the regime considered here. It is similar to its resistive counterpart described in \cite{Bierwage05b}. Both have odd parity, meaning that the magnetic islands at $\rsa$ are half a wavelength out of phase with those at $\rsb$. The slower $\Ma$-type mode has even parity (islands in phase). However, in contrast to the even-parity resistive $\Ma$-type mode \cite{Bierwage05b}, which is found in the limit of large $D_{12}$ (and eventually becomes a single tearing mode), the collisionless $\Ma$-type mode appears in the limit of $D_{12} \sim d_{\rm e}$ and peaks at both resonant surfaces. An eigenmode with such a structure has not been predicted in \cite{Mahajan82}.

\section{Nonlinear results}
\label{sec:results}

Starting from the unstable equilibrium in figure~\ref{fig:equlib_q}, all linearly unstable modes are excited by an initial perturbation of the form
\begin{equation}
\flPsi(t=0) = \frac{1}{2}\sum\limits_{m}\Psi_{0,m} r (r-1) e^{i(m\vartheta_* + \vartheta_{0,m})} + {\rm c.c.},
\label{eq:pert}
\end{equation}

\noindent where $\Psi_{0,m}$ is the perturbation amplitude (collisional case: $\Psi_{0,m} = 10^{-7}$, collisionless case: $\Psi_{0,m} = 10^{-8}$), $\vartheta_* \equiv \vartheta - \qRes^{-1}\zeta$ is a helical angle coordinate and $\vartheta_{0m}$ is an initial phase shift. The values $\vartheta_{0,m} = 0$ or $\vartheta_{0,m} = \pi$ are assigned to each $m$ in a random manner. This introduces some degree of incoherence while retaining mirror symmetry about both the $x$ and the $y$ axis (due to $\qRes = 2$ and parity conservation in RMHD). This restriction is applied for convenience and higher numerical accuracy, and has no significant effect on the phenomena discussed in this paper.

The early evolution begins with a linear phase followed by one where low-$m$ modes are nonlinearly driven by the faster high-$m$ modes. These stages were discussed in detail in \cite{Bierwage06a, Bierwage06b} and are found to be similar here. Thus, in the following, we focus on the subsequent fully nonlinear regime. The collisional case is described in section~\ref{sec:results_cdtm} and the collisionless case in section~\ref{sec:results_edtm}. The results are compared in section~\ref{sec:results_comparison}.

\subsection{Collisional case}
\label{sec:results_cdtm}

\begin{figure}
[tbp]
\centering
\includegraphics[width=13.0cm]
{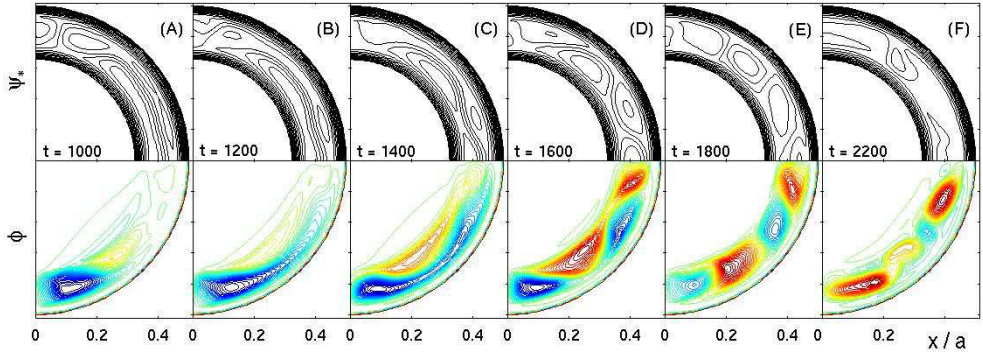}
\caption{Collisional case, $d_{\rm e} = 0$, $\SHp = 10^6$, $\RHp = 10^7$. Reconnection dynamics with $\qRes = 2$ resistive DTMs for small inter-resonance distance $D_{12} = 0.06$. The six snapshots (A)--(F) were taken during the interval $1000 \leq t \leq 2200$. Each snapshot consists of contour plots of the helical flux $\psi_* = \psi + r^2/(2\qRes)$ (top) and the electrostatic potential $\phi$ (bottom), taken in the poloidal plane at $\zeta = 0$.}
\label{fig:snaps_cdtm}%
\end{figure}

The nonlinear simulation for the collisional case was carried out for the parameter values $\SHp = 10^6$, $\RHp = 10^7$ and $d_{\rm e} = 0$. A time series of six snapshots (A)--(F), each containing contour plots of the helical flux $\psi_* = \psi + r^2/(2\qRes)$ and the electrostatic potential $\phi$, is shown in figure~\ref{fig:snaps_cdtm}. In the present case, the initial perturbation has triggered the first islands near the vertical ($y$) axis [figure~\ref{fig:snaps_cdtm}(A), top]. Their size corresponds roughly to $m=8$, although they result from a superposition of many modes. Larger islands, with $m=2$--$4$ and centered around the horizontal ($x$) axis can also be observed. However, there is no considerable $\ExB$ activity with $m=2$--$4$ in that region of the plasma [figure~\ref{fig:snaps_cdtm}(A), bottom]. This indicates that, in this stage, the larger islands constitute a yet \emph{unperturbed region}. Further evidence justifying this interpretation is presented in section~\ref{sec:results_comparison} below. In figure~\ref{fig:snaps_cdtm}(B)--(D), the localized perturbation spreads out poloidally towards the $x$-axis and breaks up the larger islands. Eventually, the whole inter-resonance region is disrupted (\emph{annular collapse}),
predominantly by a nonlinear $m=8$ DTM (D)--(F). The relaxation leads to a state with low magnetic shear in the former inter-resonance region (F).

\begin{figure}
[tb]
\centering
\includegraphics[width=13.0cm]
{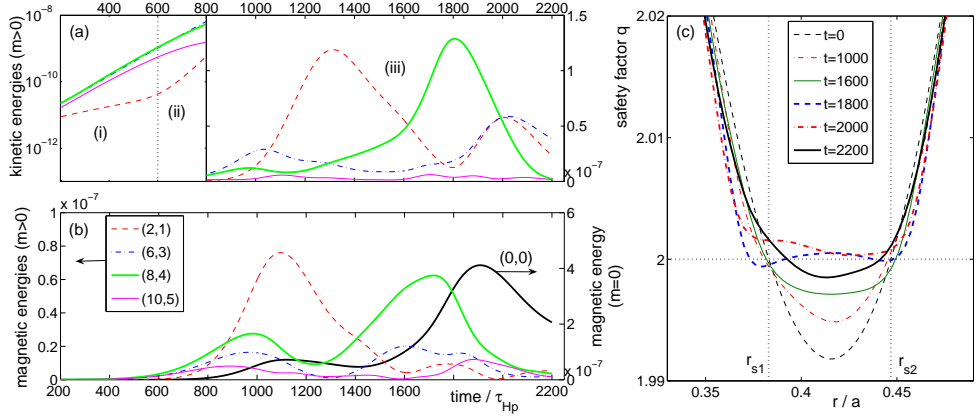}
\caption{Collisional case, $d_{\rm e} = 0$, $\SHp = 10^6$, $\RHp = 10^7$. Evolution of (a) the kinetic and (b) the magnetic energies of the modes $m=0$, $2$, $6$, $8$ and $10$. The three phases indicated in (a) are: (i) linear growth, (ii) nonlinearly driven growth of the $m=2$ mode, and (iii) annular collapse phase. (c): Evolution of the $q$ profile during the annular collapse.}
\label{fig:evol_cdtm}%
\end{figure}

The temporal evolution of the kinetic energy ($E^{\rm kin}_{m,n} = C_m\int{\rm d}r\; r|\nabla\phi_{m,n}|^2$, $C_0 = 4\pi$, $C_{m>0}=2\pi$) and the magnetic energy  ($E^{\rm mag}_{m,n} = C_m\int{\rm d}r\; r|\nabla\psi_{m,n}|^2$) is shown in figure~\ref{fig:evol_cdtm}(a) and (b) for the modes $(m,n) = (0,0)$, $(2,1)$, $(6,3)$, $(8,4)$ and $(10,5)$. Note that the profile perturbation, $m=0$ mode, has only magnetic energy, and its larger magnitude is measured on a separate axis. The labels (i) and (ii) in figure~\ref{fig:evol_cdtm}(a) indicate, respectively, the linear phase and the phase where the $m=2$ mode undergoes nonlinear driving \cite{Bierwage06b}. Note that the $m=2$ mode continues to grow beyond the stage where the driving modes saturate. This is due to the fact that $m=2$ is an unstable mode itself and the instability drive is still present at this stage, as can be inferred from the evolution of the $q$ profile in figure~\ref{fig:evol_cdtm}(c). The fully nonlinear regime begins around $t = 800$ and the label (iii) indicates the annular collapse phase mentioned in the previous paragraph. Although, the $m=2$ mode has considerable kinetic and magnetic energy during the period $1000 \lesssim t \lesssim 1400$, the contour plots in figure~\ref{fig:snaps_cdtm} show that high-$m$ islands are present at all times. The results in figure~\ref{fig:evol_cdtm}(b) suggest that the emergence of a strong $m=2$ mode has a retarding effect on the growth of the magnetic $m=8$ perturbation: $E^{\rm mag}_{8,4}$ saturates around $t=800$, has a minimum near the peak of $E^{\rm mag}_{2,1}$, and rises to its maximum ($t \approx 1700$) after the $m=2$ mode has decayed again. This is likely due to the fact that $m=2$ islands and a full chain of $m=8$ islands cannot coexist. For instance, for single tearing modes, there is experimental and numerical evidence that a large $(m,n) = (12,4)$ field has a stabilizing influence on a $(m,n)=(3,1)$ mode \cite{Wolf05}, which indicates that a large amplitude of one harmonic tends to suppress the amplitude of other harmonics on the same resonant surface. Note that the details of the dynamics seen in figure~\ref{fig:snaps_cdtm}(A)--(C) are sensitive to initial conditions, while snapshots (D)--(F) are typical for the relaxation of the present configuration.

\begin{figure}
[tbp]
\centering
\includegraphics[width=13.0cm]
{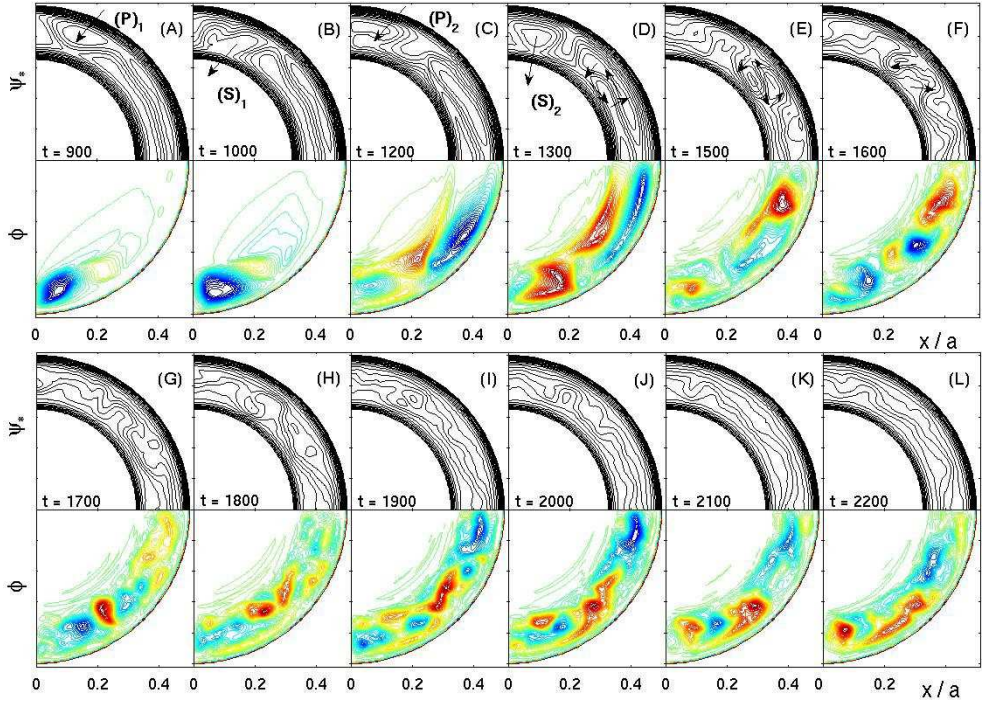}
\caption{Collisionless case, $d_{\rm e} = 0.01$, $\SHp = 10^8$, $\RHp = 10^7$. Reconnection dynamics with $\qRes = 2$ collisionless DTMs for small inter-resonance distance $D_{12} = 0.06$. The twelve snapshots (A)--(L) were taken during the interval $900 \leq t \leq 2200$. Labeled arrows in (A)--(D) indicate primary (P) and secondary reconnection (S) events [first cycle: (A) and (B); second cycle: (C) and (D)]. Arrows in (D)--(F) highlight islands revolving around each other. Otherwise arranged as figure~\protect\ref{fig:snaps_cdtm}.}
\label{fig:snaps_edtm}%
\end{figure}

The evolution of the magnetic energy of the $m=0$ mode, $E^{\rm mag}_{0,0}$ in figure~\ref{fig:evol_cdtm}(b), is closely linked to the evolution of the $q$ profile shown in figure~\ref{fig:evol_cdtm}(c). $E^{\rm mag}_{0,0}$ reaches its peak shortly after the $m=8$ mode has grown to its maximum around $t=1700$. At this point, the $m=8$ islands reach their maximal size [figure~\ref{fig:snaps_cdtm}(E)], the last magnetic surface has reconnected and the system has exhausted most of its free energy. For $t > 2000$ the energy of the profile perturbation $E^{\rm mag}_{0,0}$ decays. Correspondingly, the $q$ profile does not rise further and tends to remain close to $q \approx 2$ in the region $\rsa \lesssim r \lesssim \rsb$. $\qMin$ even drops back slightly below $\qRes = 2$. This behavior is most likely due to the resistive decay of the profile perturbation $\flPsi_{0,0}$ because the resistive time scale $\tau_{\rm R}$ for the inter-resonance region is comparable to the simulation time: $\tau_{\rm R}(D_{12})/\tHp = \hat{\eta}^{-1}\SHp (D_{12}/2)^2 \sim 10^3$. The source term $\SHp^{-1} E_0$ in equation~(\ref{eq:rmhd1}) maintains the original equilibrium profile and dissipation tends to drive the system back to the initial unstable state. The system is expected to settle down in a state where the decay of $E^{\rm mag}_{0,0}$ is balanced by weak MHD activity.

\subsection{Collisionless case}
\label{sec:results_edtm}

The nonlinear simulation for the collisionless case was carried out for the parameter values $\SHp = 10^8$, $\RHp = 10^7$ and $d_{\rm e} = 0.01$. The $d_{\rm e}$ value is just on the margin of the regime where it becomes comparable to the inter-resonance distance $D_{12}$ (cf.~figure~\ref{fig:spec_dtm}). Although the resistivity is finite, it is small enough for its effect to be negligible for both the linear instability and the prominent features of the nonlinear dynamics.

\begin{figure}
[tb]
\centering
\includegraphics[width=13cm]
{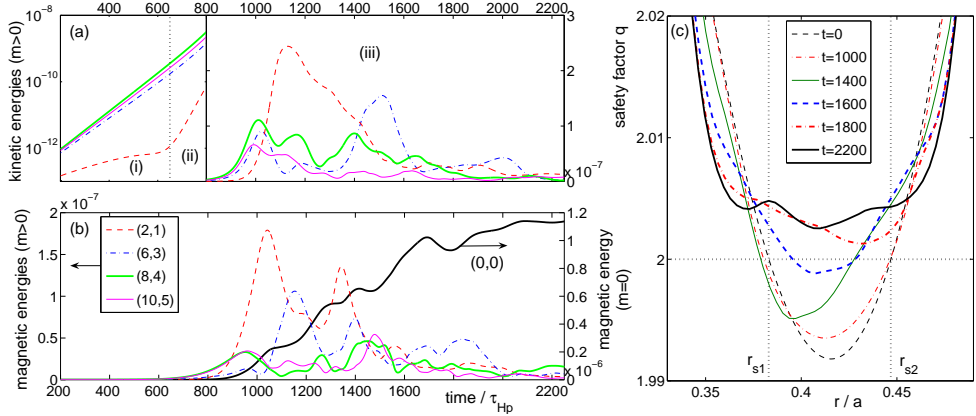}
\caption{Collisionless case, $d_{\rm e} = 0.01$, $\SHp = 10^8$, $\RHp = 10^7$. Evolution of (a) the kinetic and (b) the magnetic energies of the modes $m=0$, $2$, $6$, $8$ and $10$. The three phases indicated in (a) are: (i) linear growth, (ii) nonlinearly driven growth of the $m=2$ mode, and (iii) annular collapse phase. (c): Evolution of the $q$ profile during the annular collapse.}
\label{fig:evol_edtm}%
\end{figure}

Let us consider the sequence of twelve snapshots (A)--(L) shown in figure~\ref{fig:snaps_edtm}. As in the collisional case, $\mPeak = 8$ (cf.~figure~\ref{fig:spec_dtm}) and the same initial perturbation is used. Thus, similarly to the collisional case, the flux surfaces are first perturbed by small island structures near the $y$-axis and the yet unperturbed region takes the form of low-$m$ islands [figure~\ref{fig:snaps_edtm}(A)--(C)], which subsequently disintegrate [figure~\ref{fig:snaps_edtm}(D)--(E)]. The $m=2$ perturbation attains high magnetic and kinetic energy during the period $900 \lesssim t \lesssim 1400$, as can be observed in figure~\ref{fig:evol_edtm}(a) and (b). Nevertheless, high-$m$ islands and corresponding flows are present at all times.

In contrast to the collisional case, coherent $m=8$ islands do not form at any time. Instead, we observe increasingly turbulent structures. For instance, arrows in figure~\ref{fig:snaps_edtm}(D)--(F) indicate islands revolving around each other under the influence of an eddy. In the following snapshots, (G)--(L), the magnetic islands gradually disappear. Turbulent small-scale flows can still be observed, but with significantly reduced energies.

Close inspection of the island dynamics reveals multiple reconnection cycles. We call the process where an island forms \emph{primary reconnection} (P). During \emph{secondary reconnection} (S) the same island disappears at another location (usually on the opposite side of the inter-resonance region). In figure~\ref{fig:snaps_edtm}(A) and (B), one such reconnection cycle is indicated by arrows labeled $({\rm P})_1$ and ${\rm (S)}_1$. As can be seen in snapshots (C) and (D), the residual $\ExB$ flows in the upper part of the poloidal plane are strong enough to create another island ${\rm (P)}_2$ which is also annihilated later through secondary reconnection ${\rm (S)}_2$.

The relaxation of the $q$ profile can be observed in figure~\ref{fig:evol_edtm}(c). Note that the relaxed state has $q > 2$ everywhere. The magnetic energy of the profile perturbation, $E^{\rm mag}_{0,0}$ shown in figure~\ref{fig:evol_edtm}(b), rises relatively steadily to a level much higher than in the collisional case. Moreover, $E^{\rm mag}_{0,0}$ seems to saturate. This may again be understood in terms of the local resistive diffusion time, which is now much larger than the simulation time: $\tau_{\rm R}(D_{12})/\tHp \sim 10^5$. Although, the system considered here is only approximately collisionless, the relaxed state may be regarded as stable on the time scales of interest.

\subsection{Comparison}
\label{sec:results_comparison}

\begin{figure}
[tb]
\centering
\includegraphics[width=12.5cm]
{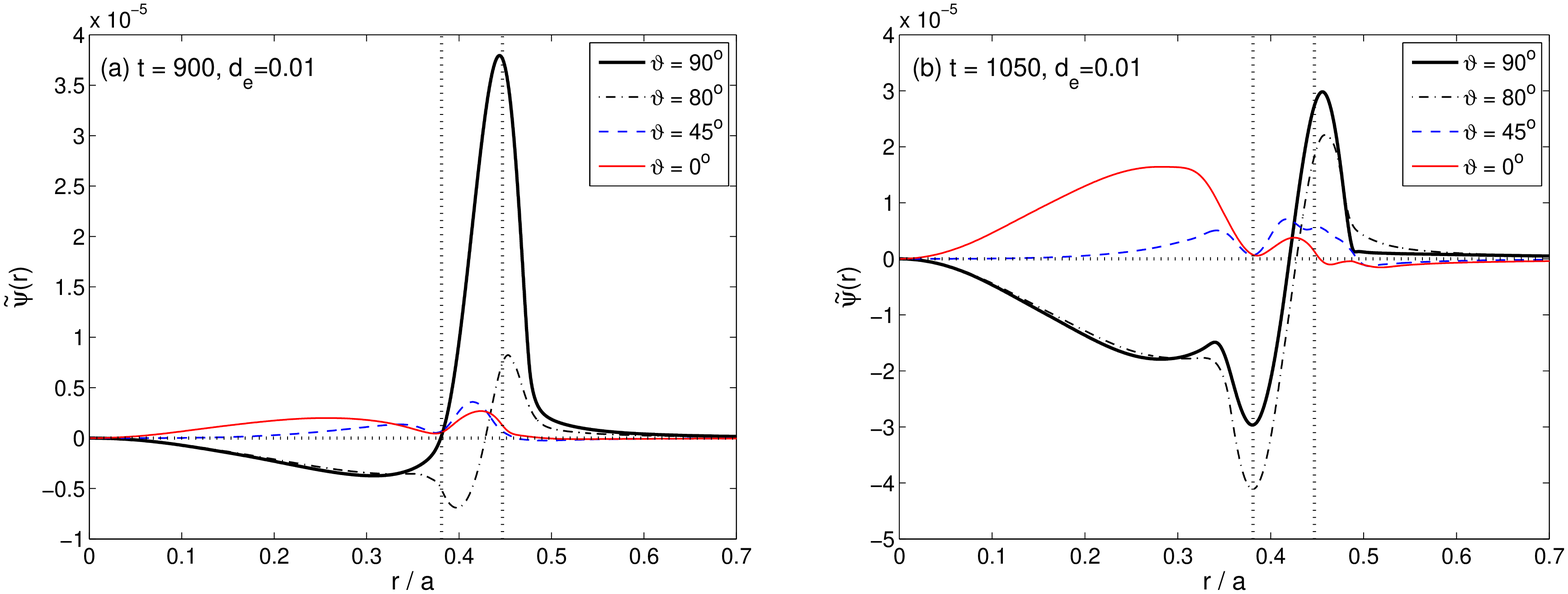}
\caption{Radial profiles of the flux perturbation $\flPsi$ in the collisionless case, measured at several poloidal angles. (a): Profiles at the same time as snapshot (a) in figure~\protect\ref{fig:snaps_edtm}. (b): Profiles at the instant where $E^{\rm mag}_{2,1}$ peaks. Vertical dotted lines indicate the original $\qRes = 2$ resonant surfaces.}
\label{fig:dpsi}%
\end{figure}

The initial response to a given initial perturbation is similar for the collisionless and collisional case. The above interpretation of the contour plots in figures~\ref{fig:snaps_cdtm} and \ref{fig:snaps_edtm}, namely that a localized perturbation is induced near the $y$-axis ($\vartheta = 90^\circ$) and that the larger islands centered at the $x$-axis ($\vartheta = 0^\circ$) represent an unperturbed region, is further confirmed by the $\flPsi$ profiles in figure~\ref{fig:dpsi}(a). The poloidal spreading of the perturbation can be seen by comparing them with the profiles in figure~\ref{fig:dpsi}(b). In figure~\ref{fig:dpsi}(b) it can be seen that even at the time where $E^{\rm mag}_{2,1}$ peaks [$t=1050$, cf.~figure~\ref{fig:evol_edtm}(b)], the perturbation near $\vartheta = 0^\circ$ is still comparatively small.

In figures~\ref{fig:evol_cdtm}(b) and \ref{fig:evol_edtm}(b), note that $E^{\rm mag}_{2,1}$ peaks \emph{before} $E^{\rm kin}_{2,1}$. This indicates that the $m=2$ mode is not a conventional DTM, but a part in a pattern resulting from interactions between several modes. The superposition of several modes with $m \sim \mPeak$ induces a localized magnetic perturbation (here, near the $y$ axis) which nonlinearly couples to low-$m$ modes (here, predominantly $m=2$) in the remaining inter-resonance region. During the peaking of the $m=2$ perturbation energy, we are thus observing a nonlinear DTM with mixed island sizes. While the detailed evolution depends on the initial conditions used in the calculation, the observed competition between different harmonics is a characteristic feature of cases where the fastest growing modes have high mode numbers $m \sim \O(10)$ and a broad-band perturbation (simulating low-amplitude background noise) is applied. This is similar for both cases studied.

Let us now compare the relaxation of the $q$ profile, which is independent of the initial conditions and thus of particular practical interest. The radial extent of the region where the $q$ profile is flattened is comparable in both cases, but the amplitude of the perturbation is larger in the collisionless case. It was shown above that one reason for this is the dissipation of the profile perturbation in the collisional case, which maintains $\qMin \approx 2$, while in the collisionless case $q > 2$ everywhere. The remaining part of this section deals with the time scales of the relaxation dynamics.

When measured in units of the poloidal Alfv\'{e}n time $\tHp$, the time intervals for nonlinear growth and decay of individual Fourier modes tend to be shorter in the collisionless case compared to the collisional case. This can be seen from the shape of the peaks of the mode energies shown in figures~\ref{fig:evol_cdtm} and \ref{fig:evol_edtm}. Nevertheless, the evolution of the $q$ profile in figures~\ref{fig:evol_cdtm}(c) and \ref{fig:evol_edtm}(c) indicates that, in both cases, the time needed to flatten the profile such that $\qMin \gtrsim 2$ is roughly $\Delta t_{\rm sat} \sim 800$ ($800 \lesssim t \lesssim 1600$). This may be explained in terms of the observation that the rapid collisionless reconnection overshoots several times before settling down. It is to be expected that the annular collapse in the collisionless case takes even more time when the damping parameters $\SHp$ and $\RHp$ are reduced. In this context, note that the formation of turbulent small-scale structures is much more pronounced in the collisionless case, despite the fact that the same value for $\RHp$ is used as in the collisional case. This suggests larger flow velocities, which is consistent with the larger kinetic energies in figure~\ref{fig:evol_edtm}(a), compared to those in figure~\ref{fig:evol_cdtm}(a).

The nonlinear saturation time in natural units is given by $T_{\rm sat} = \tHp \Delta t_{\rm sat} \propto (a/B_0) \Delta t_{\rm sat}$. In terms of system parameters, the difference between the collisional and collisionless case lies in the value of the magnetic Reynolds number $\SHp = \tau_\eta/\tA \propto a B_0 / \eta$, which is chosen here to be by a factor 100 larger in the collisionless case ($\SHp = 10^8$) compared to the collisional case ($\SHp = 10^6$). As noted above, figures~\ref{fig:evol_cdtm} and \ref{fig:evol_edtm} show that, in normalized units, the nonlinear simulation time $\Delta t_{\rm sat} \approx 800$ is similar in both cases. If we assume that the change in $\SHp$ is only due to the change of the magnetic field $B_0$ then the real relaxation time in the collisionless case is 100 times shorter than in the collisional case. If the change in $\SHp$ is assumed to be entirely due to a change in the system size $a$, then the real relaxation time is about 100 times larger in the collisionless case. Finally, if we assume that the change in $\SHp$ is only due to a change in the plasma resistivity $\eta_0$, or due to a proportional change in both $a$ and $B_0$, then the real relaxation time $T_{\rm sat} \propto (a/B_0) \Delta t_{\rm sat}$ is comparable in the two cases considered.

\section{Discussion and conclusions}
\label{sec:conclusion}

In tokamak plasmas with non-monotonic $q$ profile, pairs of nearby resonant surfaces with the same rational value $\qRes = m/n$ are produced. Examples include the current ramp-up \cite{Stix76}, current penetration after an internal disruption \cite{Kleva92}, and enhanced RS configurations where bootstrap current and external drive maintain an off-axis current density peak (e.g., \cite{Guenter00}). Motivated by the recent finding that in such configurations high-$m$ DTMs may be strongly unstable shortly after $\qMin$ drops below a low-order rational value \cite{Bierwage05b, Bierwage05a} we have analyzed the linear instability and nonlinear evolution of collisional and collisionless DTMs associated with a pair of nearby $\qRes = 2$ resonances.

A comparison between the two cases showed that both may give rise to fast growing DTMs with similar linear mode structure and high mode numbers $m \sim \O(10)$. A random broad-band perturbation was shown to induce an annular collapse involving mixed island structures, both in the collisional and collisonless case. This is in contrast to the situation typically found for large inter-resonance distances where the lowest-$m$ modes dominate linearly and produce large coherent island structures in the early stages of the nonlinear regime (e.g., \cite{White77, Ishii00}). With a broad spectrum of unstable modes, the detailed evolution depends on the initial conditions \cite{Bierwage06b}. Independently of the initial conditions, the advanced stages of the annular collapse tend to be dominated by island and $\ExB$ flow structures with $m \sim \mPeak$. The disrupted region is characterized by reduced magnetic shear and may exhibit decaying turbulent structures.

Due to the similarity of collisional and collisionless DTMs with respect to the properties mentioned above, it may be conjectured that the instability and possible dominance of high-$m$ modes in configurations with sufficiently small inter-resonance distance is a common feature of DTMs regardless of the reconnection mechanism. It should be noted that the fastest growing modes have $\mPeak \sim 10$ only if the resistivity $\eta_0 \propto \SHp^{-1}$ or the electron skin depth $d_{\rm e}$ is sufficiently large.

Differences between the collisional and the collisionless case were also identified. The nature of the reconnection mechanism does have an influence on the width of the spectrum of unstable modes. It was found that, with the same $q$ profile, many high-$m$ DTMs which are stable in the resistive case become unstable when the electron inertia effect dominates. This implies that the instability of a DTM with given mode numbers $(m,n)$ is not determined by the current profile alone, an observation which requires further investigation since it may be important for our understanding of DTM destabilization.

The reconnection and island dynamics in the collisional and collisionless case are fundamentally different from each other. After the annular collapse in the collisional case the profile perturbation decays rapidly due to the dissipative nature of the system. It is driven back towards the initial unstable state, a tendency which is balanced by continued (weak) MHD activity. In contrast, the collisionless case passes through multiple cycles of primary and secondary reconnection, during which the energy of the profile perturbation continuously rises until it saturates nonlinearly. The relaxed state is stable on the time scales of interest when $\SHp$ is chosen sufficiently large.

The results for $\qRes = 2$ DTMs are directly applicable to other values of $\qRes$ \cite{Bierwage05a}. This includes cases with nearby $\qRes = 1$ resonant surfaces for which the dynamics of resistive DTMs were recently described in \cite{Bierwage06b}. The relaxation of the $q$ profile is generally of practical interest for issues of current profile control, plasma stability and confinement.

The results presented in this paper motivate further investigations with more realistic models. To check our conjecture that for small inter-resonance distances high-$m$ DTMs may be unstable with any reconnection mechanism, it may be necessary to include finite-Larmor-radius (FLR) effects in the generalized Ohm's law (\ref{eq:ohm}) \cite{Coppi64b, Rogers96, Yu03}. This is because in a tokamak the ion sound radius $\rho_{\rm s}$ is usually larger, or at least comparable to the electron skin depth $d_{\rm e}$. Furthermore, $d_{\rm e}$ is replaced by a beta-modified natural scale length $d_{\rm s}$ \cite{Schep94}.

\ack

A.B. would like to thank S.~G\"{u}nter, S. Hamaguchi and S. Benkadda for fruitful discussions. Furthermore, he acknowledges the Max-Planck-Institut f\"{u}r Plasmaphysik Garching for its support and hospitality.

\setlength{\bibsep}{0.6pt}
\bibliographystyle{unsrt}

\end{document}